\documentstyle[12pt]{article}
\renewcommand{\theequation}{\thesection.\arabic{equation}}
\newcommand{\PP}{{\rm I} \hspace{-0.20em}{\rm P}}
\newcommand{\Q}{{\rm Q} \hspace{-0.53 em}
                 \raisebox{0.05 ex}{\rule{0.02 em}{1.33 ex}}
\hspace{0.23 em}\raisebox{0.05 ex}{\rule{0.02 em}{1.33 ex}}}
\newcommand{\LL}{{\rm I} \hspace{-0.20em}{\rm L}}
\newcommand{\GG}{{\rm G} \hspace{-0.53 em}\raisebox{0.05 ex}
           {\rule{0.02 em}{1.33 ex}}}
\newcommand{\be}{\begin{equation}}
\newcommand{\ee}{\end{equation}}
\newcommand{\bea}{\begin{eqnarray}}
\newcommand{\eea}{\end{eqnarray}}
\begin{document}
\begin{center}
{\large{\bf 
Initial-Value Problem for Inhomogeneous Condensate:\\
Gaussian Approximation and Beyond }}
\end{center}
\vspace{1.0cm}
\baselineskip=18pt
\begin{center}
Chi-Yong Lin$^{\dag}$ and A.F.R. de Toledo Piza
\end{center}
\begin{center}
{\it Instituto de F\'{\i}sica, Universidade de S\~ao Paulo \\
     Caixa Postal 66318, 05315-970, S\~ao Paulo\\
     S\~ao Paulo \ \ SP \ \ \ Brazil} 
\end{center}
\baselineskip=18pt
\vspace{1.0cm}
\begin{center}
{\bf ABSTRACT}
\end{center}
\vspace{0.5cm}

Using many-body theory we develop a set of formally exact kinetic
equations for inhomogeneous condensate and one-body observables. The method is
illutrated for $\phi^4$ field theory in 1+1 dimensions. These
equations, when computed with the help of time-dependent projection
technique, lead to a systematic mean-field expansion. The lowest and
the higher order terms correspond to, respectively, the gaussian
approximation and the dynamical correlation effect.   

\vspace{\fill}

\noindent\makebox[66mm]{\hrulefill}

\footnotesize
$^{\dag}$
Supported by Conselho Nacional de Desenvolvimento Cient\'{\i}fico e
Tecnol\'ogico (CNPq), Brazil.

\normalsize
\newpage
\baselineskip=18pt

\newpage
\begin{center}
\noindent{\bf I. Introduction}
\end{center}

The dynamical evolution of inhomogeneous field configurations is an
important problem and a common theme in cosmology, high energy 
and condensate matter physics. 
In cosmology inhomogenous field configurations appear when topological
objets such as textures or cosmic strings involving inhomogenous field
configurations are considered. Their relaxational dynamics is thought
to have a bearing on the spectrum of fluctuations in the cosmic
microwave background radiation
\cite{hindmarsh}. 
In the ultrahigh energy heavy-ion collisions ($\sqrt{s} \geq$
200 GeV/nucleon) a large energy density (few GeV/fm$^3$) is deposited
in the collision region corresponding to temperature above the
critical value for chiral symmetry restoration. In this situation, it
is possible within a volume of few fm$^3$ an inhomogeneous condensate
is formed. When these regions cool down this field relax toward the
equilibrium situation which might be misaligned with the vacuum
state
\cite{boyanovsky}. 
On the other hand, the recent sucess of the experimental observation of
Bose-Einstein condensation for systems of spin polarized magnetically
trapped alkali atoms at ultra-low tempeture is smulating development
of theory for the evolution of nonuniform condensates
\cite{stoof}.  \\
\indent A microscopic description of this
off-equilibrium process require a nonperturbative treatment of
inhomogeneous condensates and in the field-theoretical context, this
has been implemented through the use of a Gaussian ansatz for the
wavefuncctional in the framework of a time-dependent
variational principle
\cite{Ja89}. 
Actually, the Gaussian ansatz, having the form
of an exponential of quadratic form in the field operators, implies
the many-point functions, can be infact factord in terms of two-point
functions. The dynamics of the reduced two-point density becomes then
itself isoentropic, as a result of irreducible higher-order
correlation effects being neglected. 

\indent{The} purpose of the present paper is twofold. The first one is to
reevaluate and improve the gaussian approximation. We follow a
time-dependent approach developed earlier in the context of nuclear
many-body dynamics
\cite{TP82}.  
This methos allows for a formulation of a
mean-field expansion for the dynamics of the two-point correlation
function from which one recover the results of the gaussian mean-field
approximation in lowest order. Beyond this, we are able to explicitly
incluide higher dynamical correlation effects. The second purpose is to
extend the former results in the context of spatial uniformity to the
inhomogeneous field configurations
\cite{LTP92}. 
In this case the spatial
dependence of the field operator is expanded in the general natural
orbitals. These orbitals can be given in terms of an expansion of
convenient basis which willm also evolve in time according to additional
dynamical equations. Although the procedure is quite general, we will
apply our method in the context of a single scalar field in 1+1
dimensions. This will illutrate all the relevant points of the
approach and cut down inessential technical complications.

%
\medskip
\renewcommand{\theequation}{2.\arabic{equation}}
\setcounter{equation}{0}
\begin{center}
\noindent{\bf II. A Simple Example in Quantum Mechanics}
\end{center}
\smallskip

This section will illustre the scheme of approximation in the context
of $\phi^4$ in $0+1$ dimension \cite{LTP90}. Although this theory is
finite, we introduce nevertheless a counterterm for the further
use. Therefore the hamiltonian reads as 
\be
\hat H={1\over 2}\hat\pi^2
      +{m^2\over 2}\hat\phi^2
      +{g\over 24}\hat\phi^4
      -{g\over8m}\hat\phi^2
\ee
where $m$ is the renormalized mass. In the usual quantum mechanics
(QM) language, one speaks of a particle in the quartic potential with
$\phi$ and $\pi$ being its position and momentum operator  and
satisfy the usual communtation relation, $[\phi,\pi]=i$. In the
quantum field theory (QFT) language, however, one imagines that this field
lives in {\it one} point and the particles have quartic
self-interaction \cite{St84}. 

\indent A natural choice of the subsistem in the context of many-body problem
is the observables associated to the one-body density. The exact
microspic description of the time evolution of the one-body density of
a many-body system can be formally given in terms of a sum of two
parts: the usual gaussian contribution (also known as time-dependent
hartree-bogoliubov approximation) plus additional dynamics arise from
the time evolution of quantum correlations in the entire system. The
physical origin of the later contribution  lies in the complicated
dynamical evolution of quantum correlationin the entire system and
their consequences in the change in the coherence properties of each
system.

\smallskip
\indent{\bf II-a. Gaussian Variables}
\smallskip

\indent In order to derive the dynamical equations for the one-body
observables we focus on the operators which are either linear or 
bilinear forms of creation and annihilation, henceforth referred to as
gaussian observables. We begin therefore write
the Heisenberg field operators $\phi$ and $\phi$ as 
\bea
\phi(t)={1\over\sqrt{2\mu}}\Bigl[ a^{\dag}(t)+a(t) \Bigr] \hspace{1.0cm}
\pi(t) =i\sqrt{\mu\over2}\Bigl[a^{\dag}(t)-a(t) \Bigr]
\eea
where $a,a^{\dag}$ are the usual annihilation and creation operators
satisfying the boson commutation relations: $[a,a^{\dag}]=1$; the
parameter $\mu$ will be fixed later in a convenient way. The state of
the entire system is given in terms of the matrix density $F$ in the
Heisenberg picture. It is Hermitean, time indenpedent and has a unit
trace.

\indent The first variable of interest is given as mean value of
annihilation, 
\be
A_{t}={\rm Tr}a(t)F,
\ee
and its complex conjugate. From this one can define the shifted boson
operator 
\be
b=a-A.
\ee 
Next, we consider the possible mean values of the bilinear forms of
the shifted operators, which can be combined in an extended one boson
plus pairing density,
\begin{equation}
R 
\; = \; \left( \begin{array}{lc}
\langle b^{\dag}(t)b(t)\rangle \; \; & \; \; \langle b(t)b(t)\rangle \\
\langle b^{\dag}(t)b^{\dag}(t)\rangle \; \; & \; \; 
\langle b(t)b^{\dag}(t)\rangle 
\end{array}
\right)  
\; = \; \left( \begin{array}{lc}
\Lambda \; \; & \; \; \Pi \\
\Pi^{\ast} \; \; & \; \; 1 + \Lambda^{\ast} 
\end{array}
\right)  
\; = \; R^{\dag}
\end{equation} 
The quantities $\Lambda$ and $\Pi$ together with the mean-value of
field, $A$, describe the one-body observables of the system.

\indent{To} deal with the pairing density $\Pi$ we proceed, as usual,
by defining the Bogoliubov quasi-particle operator as \cite{RS80}
\be
d(t)=x_{t}^{\ast}b(t)+y^{\ast}_{t}b^{\dag}(t) \hspace{1.0cm}
d^{\dag}(t)=x_{t}b^{\dag}(t)+y_{t}b(t) 
\ee  
and require that $\langle dd\rangle = \langle d^{\dag}d^{\dag}\rangle
=0$. A sistematic way to determine the coeffcients of the Bogoliubov
transformation, $x_{t}$ and $y_{t}$, is to solve the follwoing secular
problem:
\be
G \,  R \,  X \; \; = \; \; X \, G \, N \; \; \; , 
\ee 
%
where
\be
G \; = \; 
\left( \begin{array}{cr}
1 & 0 \\ 0 & - 1
\end{array}
\right) \; \; , \;  \; \; 
X \; = \; 
\left( \begin{array}{ll}
x_{t} & y_{t}^{\ast} \\ y_{t} & x^{\ast}
\end{array}
\right) \; \;  , \; \; \; 
N \; = \; 
\left( \begin{array}{lc}
\nu_{t} & 0 \\ 0 & 1 + \nu_{t}
\end{array}
\right) \; \; \; \;  .
\ee 
The eigenvalues $\nu_{t}={\rm Tr}d^{\dag}(t)d(t)F$ stand for
occupation numbers of quasi-particles. Since the
Bogoliubov transformation is  canonica one can verify that $X_{t}$
satisfies the orthogonality and completeness relation, i.e.,
\be
X^{\dag}GX=XGX^{\dag}=G
\ee
In terms of coeficients of transformation (2.9) reads as 
\be
\mid x_{t}\mid^2-\mid y_{t}\mid^2=1
\ee
Thus, the paring densities $\langle d(t)d(t)\rangle$ and $\langle
d^{\dag}(t)d^{\dag}(t)\rangle$ can be obtained from $x_{t}$, $y_{t}$
and $\nu_{t}$. The next step is to get equations of motion for these
quantities. 

\smallskip
\indent{\bf II-b. Equation of Motion and Gaussian Approximation}
\smallskip
\\
\indent{The} Next step isto obtain the time evolution for the
variables of interest discussed in the previous subsection. We begin
with the amplitude of condensate $A_{t}$ defined by Eq.(2.3). Using
Heisenberg equation of motion one has 
\be
i\dot{A}_{t}={\rm Tr}[a,H]F=x_{t}{\rm Tr}[d,H]F+{\rm Tr}y^{\ast}[d^{\dag},H]F
\ee
where we have used (2.4) and (2.6). Now, to get equantions for the
Bogoliubov coeficients we rewrite (2.7), using (2.9), as 
\be
X^{\dag}_{t} \,  R \,  X_{t} \; = \; N . 
\ee 
Taking time derivative of this and using Eq.(2.9) we get
\be
X^{\dag}_{t} \,  \dot R \,  X_{t} \; 
= \; \dot N \, - \, \dot X^{\dag}_{t} \, R \,  X_{t} \,
            \, + \, X^{\dag}_{t} \,  R \,  \dot X_{t} \; .
\ee 
The left-hand side of this equation can be evoluated from the
Heisenberg equation of motion and (2.6) and yields
\be
i X^{\dag}_{t} \,  \dot R \,  X_{t} \; = \; 
\left( \begin{array}{cr}
{\rm Tr}[d^{\dag}d,H]F & {\rm Tr}[dd,H]F  \\ 
{\rm Tr}[d^{\dag}d^{\dag}]F & {\rm Tr}[d d^{\dag},H]F 
\end{array}
\right) \; \; , 
\ee 
The right-hand side of (2.13), on the other hand, can be obtained with
the help of (2.9). 
Equation now the result to (2.14) one gets 
%
\be
i\dot\nu= {\rm Tr}\;[d^{\dag}d,H]F 
\ee
and 
\be
i(\dot xy-x\dot y)= {\rm Tr}\;\bigl[d^{\dag}d,H\bigr]F. 
\ee
\\
\indent{The} equations (2.10), (2.15) and (2.16), together with the
normalization 
condition, determine fully, in principle, the time evolution for the
gaussian observables if the density matrix $F$ is expressible in terms
of the quantities themselves. However, when the hamiltoniana $H$
involves self-interaction fields, traces of these equations will
involve also many-body densities and therefore they are not
closed. One emergent approximation to deal with this situation is to
replace the full density $F$ by a truncated one, $F_{0}$, which has
the form of a exponential of bilinear in creation and annhilation
operators (see, e.g., Eq.(2.22) of \cite{LTP92}). 
In terms of quasi-particular operator
$F_{0}$ can be written conveniently as
\be
F_{0}(t)={1\over1+\nu_{t}}\left({\nu_{t}\over1+\nu_{t}}\right)
        ^{d^{\dag}(t)d(t)}
\ee
Notice that $F_{0}$ is diagonal in the quasi-particle basis and
contain no irreducible two or many quasi-particle correlations. 
Furthemore, it is easy to verify that $F_{0}$ given by (2.17) has 
a unit trace and reproduce the corresponding average of 
linear and bilinear field oprarors of full density, i.e.,
\bea
&&{\rm Tr}\;aF_{0}=A={\rm Tr}\;aF \nonumber\\
&&{\rm Tr}\;d^{\dag}dF_{0}=\nu={\rm Tr}\;d^{\dag}dF \nonumber\\
&&{\rm Tr}\;ddF_{0}=0={\rm Tr}\;ddF \nonumber
\eea
Therefore the approximation gives a set of self-consistent equations for the
one-body observables and is is what we refer as 
the mean-field or gaussian approximation. In
particular, when $F_{0}$ is written in terms of field representation,
it is equivalent to the density used by Jackiw \cite{Ja89}. 
This approximation
constrains the time evolution of the system to remain in a gaussian,
which contains no irreducible two or many particle correlation,
whereas the true evolution will, as time progresses, introduce
(quasi)particle-particle correlation not describable by the gaussian-like
matrix density. In terms of general discussion made in the previous
subsection the limitation of the approximation shows up in the
dynamical evolution of occupancy, 
\be
i\dot\nu \;=\; {\rm Tr}\; [d^{\dag}d,H]F_{0}
         \;=\; {\rm Tr}\; [d^{\dag}d,F_{0}]H=0.
\ee
Therefore ,  further improvements have to be achieved in order to
describe correlation effects between different subsistems.

\vspace{0.3cm}
\smallskip
\indent{\bf II-c. Projection Technique and Dynamical Correlations}
\smallskip

\indent{The} question we want to address ourselves  at this point is
how to express the full correlated density $F$ of entire system 
in terms ingredients of subsystem (gaussian variables) in such way that
one can get a set of closed equations for the one-body observable.
A framework to achieve this goal was
developed some time ago by Willis and Picard using time-dependent
projection technique \cite{WP74} in the context of master equation for
coupled systems. The method was extented later by Nemes and Piza
to study nuclear many-body dynamics \cite{TP82}. The method consists
essentially in writing  the correlation informations of the full density
in teerms of a memory kernel acting on the uncorrelated density \
$F_{0}$ \ .
 
\indent{Following} their strategy we first decompose \ $F$ \  in two parts
\begin{equation}
F \;  = \; F_0(t) + F'(t) 
\end{equation} 
where \ $F_0(t)$ \  is the uncorrelated part given by (2.17)
and therefore \ $F'(t)$ \  is a traceless correlation part. The
substitution of \ $F$ \ by just \ $F_0(t)$ \ in the equations (2.11)
and (2.15)-(2.16) 
gives the usual gaussian approximation as we have discussed before. 
The crucial step is to observe that \ $F_0(t)$ \ can 
be seen as a time-dependent projection of \ $F \, $, i.e.,
\begin{equation}
F_0 \; = \;  \PP (t) \, F \; \; \; \; \; \; \; , \; \; \; \; \; \; \; \PP (t) \, \PP (t)  \; = \; \PP (t) \; \; \; .
\end{equation} 
For the explicit construction of \ $\PP (t)$ \ we require, in addition
to eqs.~(2.20), the condition 
\begin{equation}
i \, \dot{F}_0 (t) \;  = \;  [F_0 (t) \,  , \, H] \, + \, \PP (t)[H,F] 
\end{equation} 
which is the Heisenberg  picture counterpart  of the Schr\"odinger
picture condition used in 
\cite{BFN88} 
to determine \ $\PP$ \ uniquely.  
The resulting form for  \ $\PP (t)$ \  is 
(see appendx A of Ref.\cite{LTP92} for details of the derivation)
\begin{eqnarray}
\PP \,  \cdot & = & \left\{ \left[ 1 - \; \frac{d^{\dag} \, d - p}
{1 + p}  \right] \, {\rm Tr} (\cdot) \, 
+ \,  \; \frac{d^{\dag} \, d   - p \, }{p (1 + p)} \; \,  
{\rm Tr} \left( d^{\dag} \, d \, \cdot \right)  \right. 
 +  \left[ \,  \frac{d}{p} \; \,  {\rm Tr} \left( d^{\dag} \, \cdot
  \right)
 \; + \; \frac{d^{\dag}}{1 + p }\; \, {\rm Tr} (d \, \cdot) \,  \right]
 \nonumber\\  
&+& \;  \left[ \, \frac{d \, d}{2 p \, p} \,  \; {\rm Tr} 
\left( d^{\dag} \, d^{\dag} \, \cdot \right)   \right. 
+  \left. \left. \frac{d^{\dag} \, d^{\dag}}{2(1 + p)(1 + p)} \, \;
{\rm Tr} \left( d \, d \, \cdot \right) \,  \right]  \right\} \, F_0 \; \; \; .
\end{eqnarray} 
where the dot stands for objets on which the projector acts.

\indent{The} next step  is  to obtain  a  differential  
equation for the correlated density $F'(t)$.    
This follows immediately from eqs.~(2.19) and (2.21),
\begin{equation}
\left( i \; \frac{d}{dt} \,  - \, \PP (t) \, \LL \right) \, F'(t) \; \,  = \, \; \Q \; (t) \; \LL \,  F_0(t) \; \; \; ,
\end{equation} 
\noindent{where} we have introduced the operators
\begin{equation}
\Q \; (t) \; = \; \LL - \PP (t) \; \; \; \; \; , \; \; \; \; \; \LL \, \cdot \; = \; [H \, , \, \cdot] \; \; \; .
\end{equation} 
This equation has the formal solution
\begin{equation}
F'(t) \; = \; \GG \, \; (t,0) \, F'(0) \, - \, i \int^t_0  dt' \, \GG \, \; (t,t') \, \Q \; (t') \,  \LL  \, F_0(t') \; \; \; ,
\end{equation} 
\noindent{where}  \ $\GG \, \; (t,t')$ \  is the time-ordered Green's Function
\begin{equation}
\GG \, \; (t,t') \; \,  
= \, \; T \; \exp \, i \int^t_{t'} d \tau \, \PP (\tau) \, L  \; \; \; .
\end{equation} 

\indent{What} we have obtained so far is a formaly exact expression relating
$F'(t)$ and $F_{t'}$ (for $t'\le$) and the initial correlations
$F'(0)$. This allows us to get a set of closed dynamical equations
as traces over functional of $F_{0}(t')$ and initial correlations. An
actual calculation is, however, hopeless because of the complicated
time dependence of the Heisenberg operators present in the memory
kernel of (2.25).  A systematic expansion to treat the memory 
integral of the equation~(3.8) has been discussed in 
Ref.\cite{BFN88}
in the Schr\"odinger picture. The implementation 
of the corresponding expansion in the Heisenberg picture consists 
in approximating the time evolution of the field operators, when
evoluating memory effects, by a  
simpler one-body generator of mean-field evolution, i.e.,
\be
i\dot{d}=[d,H_{0}]-i\dot{A}
+i(\dot{x}_{t}^{\ast}x_{t}-\dot{y}_{t}^{\ast}y_{t})d
-i(\dot{x}_{t}^{\ast}x_{t}^{\ast}-\dot{y}_{t}^{\ast}y_{t}^{\ast})d^{\dag}.
\ee 
The last three terms account for the (explicit) time dependence of
$d(t)$ on condensate and pairing effects; The hamiltonian $H_{0}$ is
taken, in this approximation, as a mean-field one,
\begin{eqnarray}
H_0 & = & P^{\dag} \, H +  \, d^{\dag} \; {\rm Tr}[d \, , \, H] \, F' 
- \,  d \; {\rm Tr} \left[ d^{\dag} \, , \, H \right] \, F' \nonumber \\ 
\nonumber \\
& + & \; \frac{d^{\dag} \, d^{\dag}}{2(1 + 2p)} \; 
\, {\rm Tr}[d \, d \, , \, H] \, F' 
  -  \; \frac{d \, d}{2(1 + 2p )} \; \, 
{\rm Tr} \left[d^{\dag} \, d^{\dag} \, , \, H \right] \, F' .  \; \; \; .
\end{eqnarray} 
\noindent{This} approximation results a unitary time evolution for the 
Heisenberg field operator and operators in different time are related
by a phase factor
\be
d(t)=e^{i\varphi(t,t')}d(t').
\ee
(see the third equation on page 1609 of \cite{LTP90}; see also (5.3)
of \cite{LTP92} for details). 

In this way, the authors of the 
ref.\cite{BFN88} 
devised a systematic 
expansion for the correlation density $F'(t)$ [see their Eq.(3.10)]. 
The corresponding 
expansion in the Heisenberg picture reads as
\begin{eqnarray}
F'(t) & = & \GG \, \;  (t,0) \, F'(0) \, - \, i \int^t_0  dt'_1 \, 
\Q (t_1) \, \LL  \, F_0 (t_1)  \nonumber \\ 
\nonumber \\ 
& - &  \int^t_0 dt_1 \, \left[ \, \int^t_{t_1} dt_2 \, 
\Q(t_2) \, (\LL - \LL_0 (t_2) \right] \, \Q(t_1) \, \LL \,  F_0 (t_1) 
+ \cdots  \; \; \;  ,
\end{eqnarray} 

\vspace{0.3cm}

\noindent{where} \ $\LL_0 \, \cdot = [H_0 \, , \, \cdot]$. 
\ In what follows we restrict 
ourselves to initial conditions such that \ $F'(0) = 0$ \ and to the 
lowest approximation for \ $F'(t)$. 
Therefore, the evaluation of the equations of motion 
involves traces of the type
\begin{eqnarray}
&&{\rm Tr} \, [ \hat{O} (t), H] \, F_0(t) \; 
- \; i \; {\rm Tr} \, [ \hat{O} (t), H] 
\int^t_0 dt' \, \Q \; (t') \, [H, F_0(t') ] \nonumber\\ 
&& {\rm mean-field} \hspace{2.0cm} {\rm correlation}
\end{eqnarray} 
\noindent{where} \ $\hat{O} (t)$ \  
can be \ $d(t) \, , \; d^{\dag} (t) \, d(t) \, , \; d(t) \, d(t) \, $, \ and 
operators at different times are related by equation~(2.29).

\indent{We} have now all the necessary ingredients for implementation of
the approximation and the derivation for the equation of motion is a
straightforward algebraic exercise. The results, including the
numerical calculation, for $\phi^4_{0+1}$ model are shown in
\cite{LTP90}.  
Extensions of this method to other  models of field theory were
obtained recently \cite{LTP92}. The results demonstrate that this approach
can overcome some conceptual dificults of gaussian approximation as
well as a much better description for the gaussian observables. 

%
\medskip
\medskip
\renewcommand{\theequation}{3.\arabic{equation}}
\setcounter{equation}{0}
\begin{center}
\noindent{\bf III. Kinetic Equations for Simple Observables of the Field}
\end{center}
\setcounter{section}{3}
\smallskip

\indent{Section} II reviewed some important points of our extended
gaussian approximation and its implementation in the simplest context
of quantum mechanics. In this and next section we will report its
applications in context of inhomogenoeus field configuration, which is
relevant for the dynamical evolution of many body finite system. The
simpler case of spatial uniformity was discussed for sevaral field
models recently. Technical dificulties in such cases reduce
tremendously because of translational invariance and the gaussian
variables are automatically diagonal in momentum space. In other
words, the eigenfunctions of the body-density, known as natural
orbitals, are independent of time and given by plane wave. For this
new scenary, however, the natural orbitals are time
dependent. Therefore, additional equation is needed in order to get
self-consistent equations of motion. 

\smallskip
\indent{\bf III-a. Generalized Bogoliubov Transformation}


In order to implement the idea let us first 
expand the Heisenberg field operator 
\ $\phi (t, x)$ \ and the canonical momentum \ $\pi (t, x)$ \ 
as
\begin{eqnarray}
\phi(t, x) & = & \sum_k \;  
\left[ \, f_k (x) \, a_k(t) + f^{\ast}_k (x) \, a^{\dag}_k (t) \,
\right] 
\; \; \; ,  \\
\nonumber \\ 
\pi(t, x) & = & - \, i \sum_k \;  
k_0 \left[ \, f_k (x) \, a_k(t) 
- f^{\ast}_k (x) \, a^{\dag}_k (t) \, \right] \; \; \; ,  
\end{eqnarray} 
\noindent{where} \ $a_k(t)$, \ $a_k^{\dag}(t)$ \  are 
boson operators satisfying the equal time commutation relation
\begin{equation}
[a_k (t) \, , \, a_{k'} (t)] \; = \; \delta_{kk'} \; \; \; .
\end{equation} 
The \ $f_k (x)$ \ are the periodic boundary condition plane waves
\begin{equation}
f_k (x) \; = \; \frac{{\rm e}^{i k \cdot x}}{\sqrt{2Lk_0}} \; \; \; ,
\end{equation} 
$L$ \  being  the lenght of  the periodicity box and \ $k^2_0  =  k^2 + \mu^2 \, $. \ 
The expansion mass  parameter \ $\mu$ \ is conveniently fixed, e.g. in
terms of the equilibrium solution in the mean field approximation
\cite{LTP92}.

The next step is to focus on the variables of interest, which are
mean-value of linear and bilinear boson operators.  
The first of them is the expectation value of the field operator,  
\begin{equation}
\langle \phi (t, x) \rangle \; = \; {\rm Tr} \; \phi (t, x)  \; F \; \; \; ,
\end{equation} 
where \ $F$ \ is a density matrix in the Heisenberg 
picture that characterizes the state of the system. 
In terms of the expansion (2.1), one has 
\begin{equation}
\langle \phi (t, x) \rangle \; = \; \sum_k \left[ f_k (x) \; A_k(t) + f^{\ast}_k (x) \; A^{\ast}_k(t)  \right]
\end{equation} 
with
\be
A_k (t) \; = \; {\rm Tr} \; a_k(t) \; F \; \; \; .
\ee
We can now define the shifted boson operators with the help of the \ $A_k(t)$,
\begin{equation}
b_k(t) \; = \; a_k(t) \, - \, A_k(t) \; \; \; ,
\end{equation} 
and include as variables of interest also the expectation value of pairs 
of \ $b_k(t)$, \  $b_k^{\dag} (t)$ \ at equal times: 
\begin{eqnarray}
& & \Lambda_{kk'} (t) \; = \; {\rm Tr} \, b^{\dag}_{k'} (t) \; b_k (t) \; F \; \; \; , \\
& & \Pi_{kk'} (t) \; = \; {\rm Tr} \, b_{k'} (t) \; b_k (t) \; F \; \; \; .
\end{eqnarray} 
The hermitean matrix \ $\Lambda$ \  and the symmetric matrix  \ $\Pi$ \   are in fact
 the  one-boson density  matrix and  the  pairing density  for the      shifted  bosons respectively. \ 
The corresponding matrices for the  $a$-boson are
  easily expressed  in  terms of \ $\Lambda \, $, \ $\Pi$ \ and   \ $A_k \, $. 

\indent{The} next step is to
write the one-body density matrix in diagonal form and also 
incorporate information of the pair density in an associated set of natural 
orbitals by setting up an extended one-body density matrix
\cite{RS80},
\begin{equation}
R \; = \; \left( \begin{array}{lc}
\Lambda \; \; & \; \; \Pi \\
\Pi^{\ast} \; \; & \; \; 1 + \Lambda^{\ast} 
\end{array}
\right)  \; = \; R^{\dag}
\end{equation} 
\noindent{and} solving the extendend version of eigenvalue problem 
defined by (2.7), 
where the matrix elements are given now the following matrices:
\begin{equation}
G \; = \; 
\left( \begin{array}{cr}
1 & 0 \\ 0 & - 1
\end{array}
\right) \; \; , \;  \; \; 
X \; = \; 
\left( \begin{array}{ll}
U & V^{\ast} \\ V & U^{\ast}
\end{array}
\right) \; \;  , \; \; \; 
N \; = \; 
\left( \begin{array}{lc}
P & 0 \\ 0 & 1 + P
\end{array}
\right) \; \; \; \;  .
\end{equation} 
\noindent{Since} (2.7) is a non-hermitean eigenvalue problem, 
it is important to consider also the adjoint equation
\begin{equation}
R \, G \, \tilde{X} \; \; = \; \;  \tilde{X} \, G \, N \; \; \;  \; , 
\end{equation} 
from which one finds that
\begin{equation}
\tilde{X} \; \; = \; \;  G \,  X \; \; \; \; . 
\end{equation} 
The  adjoint vectors  \ $\tilde{X}$ \  satisfy biorthogonality 
relations with \ $X$ \ 
which allow one to introduce the normalization condition
\begin{equation}
\tilde{X}^{\dag} \, X \; = \; X^{\dag} \, G \, X \; = \; G 
\end{equation} 
and the completeness relation
\begin{equation}
X \, G \, X^{\dag} \; = \;  G \; \; \; . 
\end{equation}  
\noindent{Furthermore}, one can use the eigenvectors of 
the secular problem (2.12) to construct new boson operators
\begin{equation}
\left( \begin{array}{l}
d \\ d^{\dag} 
\end{array}
  \right) \; \; = \; \; X^{\dag}  \, \left( \begin{array}{l} 
b \\ b^{\dag} \end{array}
\right) \; \; \; \; . 
\end{equation} 
\noindent{This} equation  defines in fact the general Bogoliubov 
transformation
\begin{equation}
d_{\alpha} (t) \; = \; \sum_k \, \left( U^{\ast}_{k \alpha}(t) \;
  b_k(t) \, 
+ \, V^{\ast}_{k \alpha}(t) \; b^{\dag}_k (t) \right) \; \; \; .
\end{equation} 

\noindent{It} is easy to see that the secular problem (2.12) 
constrains the expectation 
value of products of \ $d_{\alpha} (t)$,  \  $d^{\dag}_{\alpha} (t)$ \ as
\begin{eqnarray}
& & {\rm Tr} \; d^{\dag}_{\alpha} (t) \; d_{\beta} (t) \; F \; = \;  P_{\alpha} (t) \; \delta_{\alpha \beta} \\
& & {\rm Tr} \; d_{\alpha} (t) \; d_{\beta} (t) \; F \; = \;  0 \; \; \; .
\end{eqnarray} 
\noindent{where} $P_{\alpha}$ are elements of the 
eigenvaue matrix \ $P$ \ and can be 
interpreted as quasiparticle occupation numbers. 
With the help of the eq.~(2.16) one can invert the relation (2.17) as
\begin{equation}
\left( \begin{array}{l}
b \\ b^{\dag} 
\end{array}
  \right) \; \; = \; \; G \, X \, G  \, \left( \begin{array}{l} 
d \\ d^{\dag} \end{array}
\right) \; \; \; \; . 
\end{equation} 
\noindent{On} the other hand, one can now express the 
field operator in terms of the natural orbitals as
\be
\phi (t,x) \; = \; \langle \phi (t,x) \rangle \, + \, \sum_{\alpha}
\left[ v_{\alpha} (t,x) \, d_{\alpha}(t) + v^{\ast}_{\alpha} (t,x) \,
  d^{\ast}_{\alpha}(t) \right] 
\ee
with
\be
v_{\alpha} (t,x) \; = \; \sum_{k} \left[ f_k (x) \, U_{k \alpha} (t)
  \; - \; f^{\ast}_k (x) \, V_{k \alpha} (t) \right] \; \; \; .  
\ee
\noindent{Equations} have been used to obtain these results.
Therefore, our treatment for the general non-uniform field 
configuration involves 
expanding the field operator in the general natural orbitals, 
which are in turn given in terms of plane wave (or some other
convenient) expansion. 

\smallskip
\indent{\bf III-b. Formal Equations of Motion for the Gaussian
  Observables }

\indent{The} next step is to derive the equation of motion 
for these simple variables 
($A_k \,  $, \ $p_\alpha \,  $, \ $U_{k \alpha}$ \ and \ $V_{k \alpha}$). \ 
For \ $A_k(t)$ \ one finds immediately from the Heisenberg equation of motion
\begin{equation}
i \, \dot{A}_k (t) \; = \; {\rm Tr} \, [a_k (t), H] \, F  \; \; \; ,
\end{equation} 
\noindent{where} \ $H$ \ is  the field Hamiltonian. \ 
The equation of motion for the remaining  quantities can be 
obtained by taking the time-derivatives 
of the  eigenvalue equation (2.12), again in close analogy with
(2.13)-(2.14). 
%
%
%
Using the definitions for the matrices \ $N \, $,  
\ $X$ \ and \ $R \, $,  \ this equation can be written explicitly as 
\[
\left( \begin{array}{ll}
U^{\dag} \dot{\Lambda} U + U^{\dag} \dot{\Pi} V + V^{\dag} \dot{\Pi}^{\ast} U + V^{\dag} \dot{\Lambda}^{\ast} V   \; & \; 
U^{\dag} \dot{\Lambda} V^{\ast} + U^{\dag} \dot{\Pi} U^{\ast} + V^{\dag} \dot{\Pi} V^{\ast} + V^{\dag} \dot{\Lambda}^{\ast} U^{\ast} \\ 
\\
U^{\dag} \dot{\Lambda}^{\ast} V + U^{\dag} \dot{\Pi} U + V^{\dag} \dot{\Pi}^{\ast} V + V^{\dag} \dot{\Lambda} U   \; & \; 
U^{\dag} \dot{\Lambda} U^{\ast} + U^{\dag} \dot{\Pi}^{\ast} V^{\ast} + V^{\dag} \dot{\Pi} U^{\ast} + V^{\dag} \dot{\Lambda} V^{\ast}
\end{array}
 \right) 
\]

\vspace{0.2cm}

\begin{equation}
= \; 
\left( \begin{array}{ll}
\dot{P} + [U^{\dag} \dot{U} - V^{\dag} \dot{V} \, , \, P]  \; & \; 
V^{\dag} \dot{U}^{\ast} -  U^{\dag} \dot{V}^{\ast} + \{ V^{\dag} \dot{U}{\ast} -  U^{\dag} \dot{V}^{\ast} \, , \, P \}_+ \\
\\
V^{\dag} \dot{U} - U^{\dag} \dot{V} + \{ V^{\dag} \dot{U} - U^{\dag} \dot{V} \, , \, P \}_+   \; & \; 
\dot{P} + [ U^{\dag} \dot{U}^{\ast} - V^{\dag} \dot{V}^{\ast} \, , \, P ]
\end{array}
 \right)  \; \; \; ,
\end{equation} 
\noindent{where} \ $\{ \; \}_+$ \ indicates an anticommutator. \ 
Note that there are two independent matrix equations, the remaining two being 
their complex conjugates. \ The block matrix elements in the left hand side 
of eq.~(2.27) can be rewritten using the Heisenberg equation of motion as
\begin{eqnarray}
& & i \, \left[ U^{\dag} \dot{\Lambda} U + U^{\dag} \dot{\Pi} V + V^{\dag} \dot{\Pi}^{\ast} U +  V^{\dag} \dot{\Lambda}^{\ast} V \right]_{\alpha \beta} \; = \; {\rm Tr} \; \left[ d^{\dag}_{\beta} \, d_{\alpha}  \, , \, H \right] \; F \; \; \; , \; \; \; \; \\ 
\nonumber \\
& & i \, \left[ U^{\dag} \dot{\Lambda} V^{\ast} + U^{\dag} \dot{\Pi} U^{\ast} + V^{\dag} \dot{\Pi} V^{\ast} + V^{\dag} \dot{\Lambda}^{\ast} U^{\ast} \right]_{\alpha \beta} \; = \; {\rm Tr} \; \left[ d_{\beta} \, d_{\alpha} \, , \, H \right] \; F \; \; \; . \; \; \; \; 
\end{eqnarray}
\noindent{Equating} corresponding block matrix elements 
on the two sides of equation (2.27) yields
\begin{eqnarray}
& & i \, \left\{ \dot{P} + \left[ U^{\dag} \dot{U} - V^{\dag} \dot{V} \, , \,  P \right] \right\}_{\alpha \beta} \; = \; {\rm Tr} \; \left[ d^{\dag}_{\beta} \, d_{\alpha} \, , \, H \right] \; F \; \; \; , \; \; \; \; \\
\nonumber \\ 
& & i \, \left[ V^{\dag} \dot{U}^{\ast} - U^{\dag} \dot{V}^{\ast} + \left\{ V^{\dag} \dot{U}^{\ast} - U^{\dag} \dot{V}^{\ast} \, , \,  P \right\}_+ \, \right]_{\alpha \beta} \; = \; {\rm Tr} \; \left[ d_{\beta} \, d_{\alpha} \, , \, H \right] \; F \; \; \; . \; \; \; \; 
\end{eqnarray} 
\noindent{Since} \ $P$ \ is a diagonal matrix, equation~(3.28) 
splitts into two equations, 
one for \ $\alpha = \beta$ \ and other for \ $\alpha \neq \beta \, $, 
\begin{eqnarray}
& & i \, \dot{p}_{\alpha} \; = \; {\rm Tr} \; [ d^{\dag}_{\alpha} \, d_{\alpha} \, , \, H ] \; F \; \; \; , \\
\nonumber \\ 
& & i \, (p_{\beta} - p_{\alpha}) \left( U^{\dag} \dot{U} - V^{\dag} \dot{V} \right)_{\alpha \beta} \; = \; 
{\rm Tr} \; \left[ d^{\dag}_{\beta} \, d_{\alpha} \, , \, H \right] \; F  \; \; \; \; \; \; \alpha \neq \beta \; \; \; . \; \; \; \; 
\end{eqnarray}  
\noindent{Moreover}, equation (2.31) can be rewritten as
\begin{equation}
i \, \left( V^{\dag} \dot{U}^{\ast} + U^{\dag} \dot{V}^{\ast} \right)_{\alpha \beta}  \, (1 + p_{\alpha} + p_{\beta}) \; = \; 
{\rm Tr} \; \left[ d_{\beta} \, d_{\alpha} \, , \, H \right] \; F  \; \; \; .
\end{equation} 
For the particular case of a spatially uniform system, eq.~(2.33) 
is trivial since the 
plane waves are the natural orbitals.  

\indent{In} order to treat the dynamics of natural orbitals, which 
is given in 
terms of the awkward looking combinations 
$U^{\dag} \dot{U} - V^{\dag} \dot{V}$ \ and \ 
$V^{\dag} \dot{U}^{\ast} - U^{\dag} \dot{V}^{\ast}$ \ in eqs.~(2.33) and (2.34), it is convenient to first 
define matrices \ $h$ \ and \ $g$ \ as
\begin{eqnarray}
& & h \; = \; i \, \left( U^{\dag} \dot{U} \, - \, V^{\dag} \dot{V} \right) \; \; \; , \\
\nonumber \\ 
& & g \; = \; i \, \left( V^{\dag} \dot{U}^{\ast} \, - \, U^{\dag} \dot{V}^{\ast} \right) \; \; \; . 
\end{eqnarray}
\noindent{The} next step is to find simple relations between \ $\dot{U} \,  $,  \ $\dot{V}$ \ and \ $h, g \, $. \ 
The crucial point is to observe tat the dynamics of the eigenvector \ $X$ \ 
can be described as being generated by a dynamical matrix 
\ $\Omega$ \ such that 
\begin{equation}
i \, \dot{X}(t) \; = \; X(t) \; \Omega(t) \; G \; \; \; .
\end{equation} 
Using (2.16) and (2.17) one can solve eq.~(2.37) for \ $\Omega (t)$ \ as
\begin{equation}
\Omega(t) \; = \; i \, G \, X^{\dag}(t) \; G \, \dot{X} (t) \, G \; = \; 
 \left( \begin{array}{ll}
h  \; & \; g \\
g^{\ast}  \; &  \; h^{\ast} 
\end{array}
\right)  \; \; \; .
\end{equation} 
\noindent{Equating} the matrix blocks of the equation~(2.37) one finds
\begin{eqnarray}
i \, \dot{U} & = & Uh + V^{\ast} \, g^{\ast} \\
i \, \dot{V} & = & Vh + U^{\ast} \, g^{\ast} \; \; \; .
\end{eqnarray} 
Therefore, the final equations of motion for the simple variables 
of the field are eqs.~(3.24), (3.30), (3.37) and (3.38). \ 
The ingredients of the matrices \ $h$ \  and \ $g$ \ of 
eqs.~(3.37) and (3.38) are found from (3.33) and (3.34). \ 
These equations are of course not closed equations since they still
involve the full time evolution of the field operator as we have seen
in previous section. 

\indent{The} basic idea of our approximation scheme is described in
section III. The additional difficult because of the spatial
dependence of field can be handled when orbital representation is
used. The crucial point here is to notice that the reduced density 
$F_(0)(t)$ in mean-field approximation can be conveniently written as
\begin{equation}
F_0 (t) \; = \; \prod_{\alpha} \; \; 
\frac{1}{1 + p_{\alpha} (t)} \; \left[ 
\frac{p_{\alpha} (t)}{1 + p_{\alpha} (t)} \right]
^{\textstyle d^{\dag}_{\alpha} (t) \, d_{\alpha} (t)} \; \; \; \; .
\end{equation}  
\noindent{From} these ingredients one can construct the correlated density
$F'(t)$ and finally the equations of motion for a specific model.


\medskip
\smallskip
\medskip
\renewcommand{\theequation}{4.\arabic{equation}}
\setcounter{equation}{0}
\begin{center}
\noindent{\bf IV. Mean-Field and Collisional Dynamics in $\phi^4$ Field
  Theory }
\end{center}
\smallskip

\indent{Thus} far we have presented a general procedure to investigate
kinetics of one-body observable in the context of a scalar field, without
mentioning, however, any specific field model.
In this sections we will illustrate a specific example as an
application of former more formal treatment. 
We consider the $\phi^4$ Hamiltonian
\begin{equation}
H \; = \; \int dx \; {\cal H}
\end{equation} 
\begin{equation}
{\cal H} \; = \; \frac{\pi^2}{2} \, + \, \frac{1}{2} \; (\partial_x \, \phi)^2 \, + \, \frac{m^2}{2} \, \phi^2 \, + \frac{g}{4!} \, \phi^4 \, + \, \frac{\delta m^2}{2} \, \phi^2  \; \; \; .
\end{equation} 

\vspace{0.3cm}

\noindent{The} renormalization of \ $\phi^4$ \ theory in \ $1+1$ \ dimension is 
well known
\cite{stevenson} 
and can be achieved by introducing 
a mass counterterm \ $\delta m^2$ \ given as
\vspace{0.3cm}
\begin{equation}
\delta m^2 \; \; = \; \; - \; \frac{g}{4L} \, \sum_k \; \frac{1}{\sqrt{k^2 + m^2}} \; \; \; . 
\end{equation} 

\vspace{0.3cm}

The expansion of the hamiltonian on the basis of natural orbitals follows 
directly from the discussion of section~II. \  Therefore we have now all the 
necessary ingredients to implement the proposed approximation to  the 
collisional dynamics. \ This is a lengthly by straightforward algebric 
exercise. \ The resulting equations of motion are 
\vspace{0.3cm}
\begin{eqnarray*}
i \, \dot{A}_k & = & \frac{k_0}{2} \; (A_k + A^{\ast}_{-k} \, + \, \frac{m^2 + k^2}{2k_0} \; (A_k - A^{\ast}_{-k}) +  \frac{g}{24L} \;  \sum_{k_1 k_2 k_3} \; \, \frac{1}{\sqrt{k_{01} \, k_{02} \, k_{03} \, k_{0}}}   \\ 
&  &  \left \{ \left( \delta_{k_1 + k_2 + k_3 + k \, ,\, 0} \; A^{\ast}_{k_1} \, A^{\ast}_{k_2} \, A^{\ast}_{k_3} \, + \, 
 \delta_{k_1 + k_2 + k_3 - k \, , \, 0} \; A_{k_1} \, A_{k_2} \, A_{k_3} \right) \right.  \\ 
&  &  3 \, \left. \left( \delta_{k_1 + k_2 - k_3 + k \, ,\, 0} \; A^{\ast}_{k_1} \, A^{\ast}_{k_2} \, A_{k_3} \, + \, 
 \delta_{k_1 + k_2 - k_3 - k \, , \, 0} \; A_{k_1} \, A_{k_2} \, A^{\ast}_{k_3} \right) \right \}  \\ 
& + & \frac{g}{8L} \;  \sum_{k_1 k_2 k_3} \; \, \frac{1}{\sqrt{k_{01} \, k_{02} \,k_{03} \, k_{0}}}   \\ 
&  &  \left( \delta_{k_1 + k_2 - k_3  \, ,\, 0} \; A^{\ast}_{k_1} \, + \,  \delta_{k_1 - k - k_2 + k_3  \, , \, 0} \; A_{k_1} \right)  \; \sum_{\alpha} \; T_{k_2 \alpha} \, T^{\ast}_{k_3 \alpha} \, (1 + 2p_{\alpha})  \\ 
& - & \frac{g}{8L k_0} \; (A_k + A^{\ast}_{-k} \, \sum_{k'} \; \frac{1}{\sqrt{k'^{2} + m^2}} \, + \, i \, \Gamma_A (t) \; \; \; ,  
\end{eqnarray*} 
\begin{equation}
\dot{p}_{\alpha} \; \; = \; \; \Gamma_p (t) \; \; \; ,
\end{equation} 

\begin{eqnarray}
\lefteqn{(1 + p_\alpha + p_\beta ) \, g_{\alpha \beta}} \nonumber \\ 
& = &  \sum_k \, \left[ - \; \frac{k_0}{2} \; (U^{\ast}_{k \alpha} + V^{\ast}_{- k \alpha} ) (U^{\ast}_{- k \beta} + V^{\ast}_{ k \beta} ) \,  + \, \frac{k^2 + m^2}{2k_0} \; 
(U^{\ast}_{k \alpha} - V^{\ast}_{- k \alpha} ) (U^{\ast}_{- k \beta} - V^{\ast}_{ k \beta} )  \right] \nonumber \\ 
& + &  \frac{g}{8L} \;  \sum_{k_1 k_2 k_3 k_4} \; \, \frac{1 + p_{\alpha} + p_{\beta}}{\sqrt{k_{01} \, k_{02} \, k_{03} \, k_{04}}}  \; 
 \left\{ \left( \delta_{k_1 + k_2 + k_3 + k_4 \, ,\, 0} \; A^{\ast}_{k_1} \, A^{\ast}_{k_2}  \, + \, 
 \delta_{k_1 + k_2 - k_3 - k_4 \, , \, 0} \; A_{k_1} \, A_{k_2} \right. \right.  \nonumber \\ 
& + &  \left. \left. 2 \delta_{k_1 - k_2 + k_3 + k_4 \, ,\,  0} \; A^{\ast}_{k_1} \, A_{k_2} 
 \right) \; T^{\ast}_{k_3 \, \alpha} \, T_{k_4 \, \beta} \right\} \nonumber \\ 
& + &    \frac{g}{8L} \;  \sum_{k_1 k_2 k_3 k_4} \; \delta_{k_1 + k_2 + k_3 - k_4 \, ,\, 0} \; \, \frac{1 + p_{\alpha} + p_{\beta}}{\sqrt{k_{01} \, k_{02} \, k_{03} \, k_{04}}} \; \, T^{\ast}_{k_1 \, \alpha} \, T^{\ast}_{k_2 \, \beta} \, \sum_{\alpha'} \; T^{\ast}_{k_3 \, \alpha'} \, T_{k_4 \, \alpha'} \, (1 + 2 p_{\alpha'} ) \nonumber \\ 
& - & \frac{g}{8L} \;  \sum_{k_1} \; \frac{1 + p_{\alpha} + p_{\beta}}{k_{01}} \; \, T^{\ast}_{k_1 \, \alpha} \, T^{\ast}_{k_2 \, \beta} \, \sum_{k_2} \; \frac{1}{\sqrt{k^2_2 + m^2}} \, - \, i \, \Gamma_g \, (t) \; \; \; . \\  
\nonumber \\ 
\lefteqn{(p_\beta - p_\alpha) \, h_{\alpha \beta}} \nonumber \\ 
& = &  \sum_k \, \left[ \frac{k_0}{2} \; (U^{\ast}_{k \alpha} + V^{\ast}_{- k \alpha} ) (U_{k \beta} + V_{- k \beta} ) + \frac{k^2 + m^2}{2k_0} \; \, 
T^{\ast}_{k \alpha} \,  T_{k \beta}  \right] (p_\beta - p_\alpha) \nonumber \\ 
& + &  \; \frac{g}{16L} \;  \sum_{k_1 k_2 k_3 k_4} \; \, \frac{1}{\sqrt{k_{01} \, k_{02} \, k_{03} \, k_{04}}}  \nonumber \\ 
&  &   \left\{ \delta_{k_1 + k_2 + k_3 - k_4 \, ,\, 0} \; \left( A^{\ast}_{k_1} \, A^{\ast}_{k_2}  \, T^{\ast}_{k_3 \alpha} \, T_{k_4 \beta} \, + \, A_{k_1} \, A_{k_2}  \, T_{k_3 \beta} \, T^{\ast}_{k_4 \alpha} \right) \right.  \nonumber \\
& + &   \delta_{k_1 + k_2 - k_3 + k_4 \, ,\, 0} \; \left( A_{k_1} \, A_{k_2}  \, T^{\ast}_{k_3 \alpha} \, T_{k_4 \beta} \, + \, A^{\ast}_{k_1} \, A^{\ast}_{k_2}  \, T_{k_3 \beta} \, T^{\ast}_{k_4 \alpha} \right)  \nonumber \\
& + &  \left. 2 \delta_{k_1 - k_2 + k_3 - k_4 \, ,\, 0} \; \left( A^{\ast}_{k_1} \, A_{k_2}  \, T^{\ast}_{k_3 \alpha} \, T_{k_4 \beta} \, + \, A_{k_1} \, A^{\ast}_{k_2}  \, T_{k_3 \beta} \, T^{\ast}_{k_4 \alpha} \right) \right \} 
\; (p_{\beta} - p_{\alpha})  \nonumber \\ 
& + &    \frac{g}{8L} \;  \sum_{k_1 k_2 k_3 k_4} \; \frac{\delta_{k_1 + k_2 - k_3 - k_4 \, ,\, 0}}{\sqrt{k_{01} \, k_{02} \, k_{03} \, k_{04}}} \; \, T^{\ast}_{k_1 \, \alpha} \, T_{k_3 \, \beta} \, \sum_{\alpha'} \; T^{\ast}_{k_2 \, \alpha'} \, T_{k_4 \, \alpha'} \, (p_{\beta} -  p_{\alpha} ) \nonumber \\ 
& - &   \frac{g}{8L} \;  \sum_{k_1} \; \frac{1}{k_{01}} \; \, T^{\ast}_{k_1 \, \alpha} \, T_{k_1 \, \beta} \, \sum_{k_2} \; \frac{1}{\sqrt{k^2_2 + m^2}} \; (p_{\beta} - p_{\alpha}) + i \, \Gamma_h \, (t) \; \; \; ,
\end{eqnarray} 

\vspace{0.3cm}

where 
\[
T_{k \alpha} \; \; = \; \; U_{k \alpha} \, - \, V_{k \alpha}
\]

\vspace{0.3cm}

and the collision integrals \ $\Gamma (t)$ \ are
\vspace{0.3cm}
\begin{eqnarray}
 \Gamma_A (t) & = & 
\frac{g^2}{96L^2} \;  \sum_{k_1 k_2 k_3} \; \, \frac{1}{\sqrt{k_{01} \, k_{02} \, k_{03} \, k_{0}}}  \; \, \sum_{\alpha_1 \alpha_2 \alpha_3} \nonumber \\ 
&  &  \left\{ \delta_{k_1 + k_2 + k_3 + k \, ,\, 0} \;  T^{\ast}_{k_1 \alpha_1} \, T^{\ast}_{k_2 \alpha_2} \, T^{\ast}_{k_3 \alpha_3} \, I^{(4)}_{\alpha_1 \alpha_2 \alpha_3}  \right. \nonumber \\
& - &   \delta_{k_1 + k_2 + k_3 - k \, ,\, 0} \;  T_{k_1 \alpha_1} \, T_{k_2 \alpha_2} \, T_{k_3 \alpha_3} \, I^{(4)^{\ast}}_{\alpha_1 \alpha_2 \alpha_3} \,  \nonumber \\
& + & 3  \delta_{k_1 + k_2 + k - k_3 \, ,\, 0} \;  T^{\ast}_{k_1 \alpha_1} \, T^{\ast}_{k_2 \alpha_2} \, T_{k_3 \alpha_3} \, I^{(5)}_{\alpha_1 \alpha_2 \alpha_3}  \nonumber \\
& - &  \left. 3 \delta_{k_1 + k_2 - k -  k_3 \, ,\, 0} \;  T_{k_1 \alpha_1} \, T_{k_2 \alpha_2} \, T^{\ast}_{k_3 \alpha_3} \, I^{(5)^{\ast}}_{\alpha_1 \alpha_2 \alpha_3} \,  \right\} \; \; \; , \\  
\nonumber \\ 
 \Gamma_p (t) & = & 
\frac{g^2}{96L^2} \;  \sum_{k_1 k_2 k_3 k_4} \; \, \frac{1}{\sqrt{k_{01} \, k_{02} \, k_{03} \, k_{04}}}  \; \, \sum_{\alpha_1 \alpha_2 \alpha_3} \nonumber \\ 
&  &  \left\{ \delta_{k_1 + k_2 + k_3 + k_4 \, ,\, 0} \;  T^{\ast}_{k_1 \alpha_1} \, T^{\ast}_{k_2 \alpha_2} \, T^{\ast}_{k_3 \alpha_3} \, T^{\ast}_{k_4 \alpha} \, I^{(1)}_{\alpha_1 \alpha_2 \alpha_3 \alpha}  \right. \nonumber \\
& + &  3 \delta_{k_1 + k_2 + k_3 - k_4 \, ,\, 0} \;  T^{\ast}_{k_1 \alpha_1} \, T^{\ast}_{k_2 \alpha_2} \, T^{\ast}_{k_3 \alpha} \, T_{k_4 \alpha_3} \, I^{(2)}_{\alpha_1 \alpha_2 \alpha \alpha_3} \nonumber \\
& - &  \delta_{k_1 + k_2 + k_3 - k_4 \, ,\, 0} \;  T^{\ast}_{k_1 \alpha_1} \, T^{\ast}_{k_2 \alpha_2} \, T^{\ast}_{k_3 \alpha_3} \, T_{k_4 \alpha} \, I^{(2)}_{\alpha_1 \alpha_2 \alpha_3 \alpha}  \nonumber \\
& + &  \left. 3 \delta_{k_1 + k_2 - k_3 -  k_4 \, ,\, 0} \;  T^{\ast}_{k_1 \alpha} \, T^{\ast}_{k_2 \alpha_1} \, T_{k_3 \alpha_2} \, T_{k_4 \alpha_3} \, I^{(3)}_{\alpha \alpha_1 \alpha_2 \alpha_3} \,  \right\}   \nonumber \\
& + & \frac{g^2}{32L^2} \;  \sum_{k_1 k_2 k_3 k_4} \; \, \frac{1}{\sqrt{k_{01} \, k_{02} \, k_{03} \, k_{04}}}  \; \, \sum_{\alpha_1 \alpha_2} \nonumber \\ 
&  &  \left\{ \left( \delta_{k_1 + k_2 + k_3 + k_4 \, ,\, 0} \, A_{k_1}  + 
\delta_{- k_1 + k_2 + k_3 + k_4 \, ,\, 0} \, A^{\ast}_{k_1} \right) 
\;  T_{k_2 \alpha_1} \, T_{k_3 \alpha_2} \,  T_{k_4 \alpha} \, I^{(4)}_{\alpha_1 \alpha_2  \alpha}  \right. \nonumber \\
&  & 2 \left( \delta_{k_1 - k_2 - k_3 + k_4 \, ,\, 0} \, A_{k_1}  + 
\delta_{k_1 + k_2 + k_3 - k_4 \, ,\, 0} \, A^{\ast}_{k_1} \right) 
\;  T^{\ast}_{k_2 \alpha_1} \, T^{\ast}_{k_3 \alpha} \,  T_{k_4 \alpha_2} \, I^{(5)}_{\alpha_1 \alpha  \alpha_2}   \nonumber \\
& - &  \left. \left( \delta_{k_1 - k_2 - k_3 + k_4 \, ,\, 0} \, A_{k_1}  + 
\delta_{k_1 + k_2 + k_3 - k_4 \, ,\, 0} \, A^{\ast}_{k_1} \right) 
\;  T^{\ast}_{k_2 \alpha_1} \, T^{\ast}_{k_3 \alpha_2} \,  T_{k_4 \alpha} \, I^{(5)}_{\alpha_1 \alpha_2  \alpha} \right\} \nonumber \\
& + & {\rm c. \, c.} \; \; \; \; ,  \\  
\nonumber \\ 
 \Gamma_h (t) & = & 
\frac{g^2}{96L^2} \;  \sum_{k_1 k_2 k_3 k_4} \; \, \frac{1}{\sqrt{k_{01} \, k_{02} \, k_{03} \, k_{04}}}  \; \, \sum_{\alpha_1 \alpha_2 \alpha_3} \nonumber \\ 
& &  \left\{ \delta_{k_1 + k_2 + k_3 + k_4 \, ,\, 0} \;  T^{\ast}_{k_1 \alpha_1} \, T^{\ast}_{k_2 \alpha_2} \, T^{\ast}_{k_3 \alpha_3} \, T^{\ast}_{k_4 \alpha} \, I^{(1)}_{\alpha_1 \alpha_2 \alpha_3 \beta}  \right. \nonumber \\
& + &  3 \delta_{k_1 + k_2 + k_3 - k_4 \, ,\, 0} \;  T^{\ast}_{k_1 \alpha_1} \, T^{\ast}_{k_2 \alpha_2} \, T^{\ast}_{k_3 \alpha} \, T_{k_4 \alpha_3} \, I^{(2)}_{\alpha_1 \alpha_2 \beta \alpha_3} \nonumber \\
& - &  \delta_{k_1 + k_2 + k_3 - k_4 \, ,\, 0} \;  T_{k_1 \alpha_1} \, T_{k_2 \alpha_2} \, T_{k_3 \alpha_3} \, T^{\ast}_{k_4 \alpha} \, I^{(2)^{\ast}}_{\alpha_1 \alpha_2 \alpha_3 \beta}  \nonumber \\
& + &  \left. 3 \delta_{k_1 + k_2 - k_3 -  k_4 \, ,\, 0} \;  T^{\ast}_{k_1 \alpha} \, T^{\ast}_{k_2 \alpha_1} \, T_{k_3 \alpha_2} \, T_{k_4 \alpha_3} \, I^{(3)}_{\beta \alpha_1 \alpha_2 \alpha_3} \,  \right\}   \nonumber \\
& + & \frac{g^2}{32L^2} \;  \sum_{k_1 k_2 k_3 k_4} \; \, \frac{1}{\sqrt{k_{01} \, k_{02} \, k_{03} \, k_{04}}}  \; \, \sum_{\alpha_1 \alpha_2} \nonumber \\ 
& &  \left\{ \left( \delta_{- k_1 + k_2 + k_3 + k_4 \, ,\, 0} \, A_{k_1}  + 
\delta_{k_1 + k_2 + k_3 + k_4 \, ,\, 0} \, A^{\ast}_{k_1} \right) 
\;  T^{\ast}_{k_2 \alpha_1} \, T^{\ast}_{k_3 \alpha_2} \,  T^{\ast}_{k_4 \alpha} \, I^{(4)}_{\alpha_1 \alpha_2  \beta}  \right. \nonumber \\
& & 2 \left( \delta_{k_1 - k_2 - k_3 + k_4 \, ,\, 0} \, A_{k_1}  + 
\delta_{k_1 + k_2 + k_3 - k_4 \, ,\, 0} \, A^{\ast}_{k_1} \right) 
\;  T^{\ast}_{k_2 \alpha_1} \, T^{\ast}_{k_3 \alpha} \,  T_{k_4 \alpha_2} \, I^{(5)}_{\alpha_1 \beta \alpha_2}   \nonumber \\
& - &  \left. \left( \delta_{k_1 - k_2 - k_3 + k_4 \, ,\, 0} \, A_{k_1}  + 
\delta_{k_1 + k_2 + k_3 - k_4 \, ,\, 0} \, A^{\ast}_{k_1} \right) 
\;  T_{k_2 \alpha_1} \, T_{k_3 \alpha_2} \,  T^{\ast}_{k_4 \alpha} \, I^{(5)}_{\alpha_1 \alpha_2  \beta} \right\} \nonumber \\
& + & {\rm c. \, c.} \; (\alpha \leftrightarrow \beta)\; \; \; \; , \\  
\nonumber \\ 
\Gamma_g (t) & = & 
\frac{g^2}{96L^2} \;  \sum_{k_1 k_2 k_3 k_4} \; \, \frac{1}{\sqrt{k_{01} \, k_{02} \, k_{03} \, k_{04}}}  \; \, \sum_{\alpha_1 \alpha_2 \alpha_3} \nonumber \\ 
&  &  \left\{ \delta_{k_1 + k_2 + k_3 - k_4 \, ,\, 0} \;  T_{k_1 \alpha_1} \, T_{k_2 \alpha_2} \, T_{k_3 \alpha_3} \, T^{\ast}_{k_4 \alpha} \, I^{(1)^{\ast}}_{\alpha_1 \alpha_2 \alpha_3 \beta}  \right. \nonumber \\
& + &  3 \delta_{k_1 + k_2 - k_3 - k_4 \, ,\, 0} \;  T_{k_1 \alpha_1} \, T_{k_2 \alpha_2} \, T^{\ast}_{k_3 \alpha} \, T^{\ast}_{k_4 \alpha_3} \, I^{(2)^{\ast}}_{\alpha_1 \alpha_2 \beta \alpha_3} \nonumber \\
& - &  \delta_{k_1 + k_2 + k_3 + k_4 \, ,\, 0} \;  T^{\ast}_{k_1 \alpha_1} \, T^{\ast}_{k_2 \alpha_2} \, T^{\ast}_{k_3 \alpha_3} \, T^{\ast}_{k_4 \alpha} \, I^{(2)}_{\alpha_1 \alpha_2 \alpha_3 \beta}  \nonumber \\
& + &  \left. 3 \delta_{k_1 + k_2 + k_3 -  k_4 \, ,\, 0} \;  T^{\ast}_{k_1 \alpha_1} \, T^{\ast}_{k_2 \alpha_2} \, T^{\ast}_{k_3 \alpha} \, T_{k_4 \alpha_3} \, I^{(3)}_{\alpha_1 \alpha_2 \beta \alpha_3} \,  \right\}   \nonumber \\
& + & \frac{g^2}{32L^2} \;  \sum_{k_1 k_2 k_3 k_4} \; \, \frac{1}{\sqrt{k_{01} \, k_{02} \, k_{03} \, k_{04}}}  \; \, \sum_{\alpha_1 \alpha_2} \nonumber \\ 
&  &  \left\{ \left( \delta_{k_1 + k_2 + k_3 - k_4 \, ,\, 0} \, A_{k_1}  + 
\delta_{k_1 - k_2 - k_3 + k_4 \, ,\, 0} \, A^{\ast}_{k_1} \right) 
\;  T_{k_2 \alpha_1} \, T_{k_3 \alpha_2} \,  T^{\ast}_{k_4 \alpha} \, I^{(4)^{\ast}}_{\alpha_1 \alpha_2  \beta}  \right. \nonumber \\
&  & 2 \left( \delta_{k_1 - k_2 - k_3 + k_4 \, ,\, 0} \, A_{k_1}  + 
\delta_{k_1 + k_2 + k_3 - k_4 \, ,\, 0} \, A^{\ast}_{k_1} \right) 
\;  T_{k_2 \alpha_1} \, T^{\ast}_{k_3 \alpha} \,  T^{\ast}_{k_4 \alpha_2} \, I^{(5)^{\ast}}_{\alpha_1 \beta \alpha_2}   \nonumber \\
& - &  \left. \left( \delta_{- k_1 + k_2 + k_3 + k_4 \, ,\, 0} \, A_{k_1}  + 
\delta_{k_1 + k_2 + k_3 + k_4 \, ,\, 0} \, A^{\ast}_{k_1} \right) 
\;  T^{\ast}_{k_2 \alpha_1} \, T^{\ast}_{k_3 \alpha_2} \,  T^{\ast}_{k_4 \alpha} \, I^{(5)}_{\alpha_1 \alpha_2  \beta} \right\} \nonumber \\
& + &  \alpha \; \leftrightarrow \; \beta\; \; \; \; . 
\end{eqnarray} 

\vspace{0.3cm}

In these equations we used the abbreviations 
\vspace{0.3cm}
\begin{eqnarray}
I^{(1)}_{\alpha_1 \alpha_2 \alpha_3 \alpha_4}  (t) & = & 
\int^t_0 dt' \; \sum_{k'_1 k'_2 k'_3 k'_4} \; \, \frac{\delta_{k'_1 +  k'_2 + k'_3 + k'_4 \, , \, 0}}{\sqrt{k'_{01} \, k'_{02} \, k'_{03} \, k'_{04}}} \, \; \sum_{\gamma_1 \gamma_2 \gamma_3 \gamma_4} \; \left( T_{k'_1 \gamma_1} \, T_{k'_2 \gamma_2} \, T_{k'_3 \gamma_3} \, T_{k'_4 \gamma_4}  \right)_{t'} \nonumber \\
&  & \left( 1 \, + \, \sum^4_i \; p_{\gamma_i} \, + \, \sum^4_{i < j} \; p_{\gamma_i} \, p_{\gamma_j} \, + \, 
\sum^4_{i < j < \ell} \; p_{\gamma_i} \, p_{\gamma_j} \, p_{\gamma_{\ell}} \right)_{t'} \nonumber \\
&  & \left( M^{\ast}_{\alpha_1 \gamma_1} (t,t') \, 
M^{\ast}_{\alpha_2 \gamma_2} (t,t') \, 
M^{\ast}_{\alpha_3 \gamma_3} (t,t') \, 
M^{\ast}_{\alpha_4 \gamma_4} (t,t')   \right) \; \; \; , \\  
\nonumber \\ 
I^{(2)}_{\alpha_1 \alpha_2 \alpha_3 \alpha_4}  (t) & = & 
\int^t_0 dt' \; \sum_{k'_1 k'_2 k'_3 k'_4} \; \, \frac{\delta_{k'_1 +  k'_2 + k'_3 - k'_4 \, , \, 0}}{\sqrt{k'_{01} \, k'_{02} \, k'_{03} \, k'_{04}}} \, \; \sum_{\gamma_1 \gamma_2 \gamma_3 \gamma_4} \; \left( T_{k'_1 \gamma_1} \, T_{k'_2 \gamma_2} \, T_{k'_3 \gamma_3} \, T^{\ast}_{k'_4 \gamma_4}  \right)_{t'} \nonumber \\
&  & \left( p_{\gamma_4} \, - \,  p_{\gamma_1} \, p_{\gamma_2} \, p_{\gamma_3} \, + \, p_{\gamma_4} \; \sum^3_i \; p_{\gamma_i}  \, + \, 
p_{\gamma_4} \; \sum^3_{i < j} \; p_{\gamma_i} \, p_{\gamma_j}  \right)_{t'} \nonumber \\
&  & \left( M^{\ast}_{\alpha_1 \gamma_1} (t,t') \, 
M^{\ast}_{\alpha_2 \gamma_2} (t,t') \, 
M^{\ast}_{\alpha_3 \gamma_3} (t,t') \, 
M_{\alpha_4 \gamma_4} (t,t') \,   \right) \; \; \; ,   
\end{eqnarray} 


\begin{eqnarray}
I^{(3)}_{\alpha_1 \alpha_2 \alpha_3 \alpha_4}  (t) & = & 
\int^t_0 dt' \; \sum_{k'_1 k'_2 k'_3 k'_4} \; \, \frac{\delta_{k'_1 +  k'_2 - k'_3 - k'_4 \, , \, 0}}{\sqrt{k'_{01} \, k'_{02} \, k'_{03} \, k'_{04}}} \, \; 
\sum_{\gamma_1 \gamma_2 \gamma_3 \gamma_4} \; \left( T_{k'_1 \gamma_1} \, T_{k'_2 \gamma_2} \, T^{\ast}_{k'_3 \gamma_3} \, T^{\ast}_{k'_4 \gamma_4}  \right)_{t'} \nonumber \\
&  & \left[ p_{\gamma_3} \, p_{\gamma_4} (1 + p_{\gamma_1}) (1 + p_{\gamma_2}) \, - \, p_{\gamma_1} \, p_{\gamma_2} (1 + p_{\gamma_3})(1 + p_{\gamma_4})  \right]_{t'}
 \nonumber \\
& & \left( M^{\ast}_{\alpha_1 \gamma_1} (t,t') \, 
M^{\ast}_{\alpha_2 \gamma_2} (t,t') \, 
M_{\alpha_3 \gamma_3} (t,t') \, 
M_{\alpha_4 \gamma_4} (t,t') \,   \right) \; \; \; , \\ 
\nonumber \\ 
I^{(4)}_{\alpha_1 \alpha_2 \alpha_3}  (t) & = & 
\int^t_0 dt' \; \sum_{k'_1 k'_2 k'_3 k'_4} \; \, \frac{1}{\sqrt{k'_{01} \, k'_{02} \, k'_{03} \, k'_{04}}} \, 
 \; \sum_{\gamma_1 \gamma_2 \gamma_3} \; \left( T_{k'_2 \gamma_1} \, T_{k'_3 \gamma_2} \, T_{k'_4 \gamma_3}  \right)_{t'} \nonumber \\
&  & \left( \delta_{k'_1 +  k'_2 + k'_3 + k'_4 \, , \, 0} \, A_{k'_1} \, + \, 
\delta_{- k'_1 +  k'_2 + k'_3 + k'_4 \, , \, 0} \, A^{\ast}_{k'_1} \right)_{t'} \nonumber \\
&  & \left( 1 \, + \,  \sum^3_{i} \; p_{\gamma_i} \, + \, 
\sum^3_{i < j} \; p_{\gamma_i} \, p_{\gamma_j} 
  \right)_{t'} \nonumber \\
&  & \left( M^{\ast}_{\alpha_1 \gamma_1} (t,t') \, 
M^{\ast}_{\alpha_2 \gamma_2} (t,t') \, 
M^{\ast}_{\alpha_3 \gamma_3} (t,t')  
  \right) \; \; \; , \\  
\nonumber \\ 
I^{(5)}_{\alpha_1 \alpha_2 \alpha_3}  (t) & = & 
\int^t_0 dt' \; \sum_{k'_1 k'_2 k'_3 k'_4} \; \, \frac{1}{\sqrt{k'_{01} \, k'_{02} \, k'_{03} \, k'_{04}}} \, 
 \; \sum_{\gamma_1 \gamma_2 \gamma_3} \; \left( T_{k'_2 \gamma_1} \, T_{k'_3 \gamma_2} \, T^{\ast}_{k'_4 \gamma_3}  \right)_{t'} \nonumber \\
&  & \left( \delta_{k'_1 +  k'_2 + k'_3 - k'_4 \, , \, 0} \, A_{k'_1} \, + \, 
\delta_{k'_1 - k'_2 - k'_3 + k'_4 \, , \, 0} \, A^{\ast}_{k'_1} \right)_{t'} \nonumber \\
&  & \left[ \, p_{\gamma_3} \,  (1 + p_{\gamma_1} +  p_{\gamma_2}) \, - \, p_{\gamma_1} \, p_{\gamma_2} \right]_{t'} \nonumber \\
&  & \left( M^{\ast}_{\alpha_1 \gamma_1} (t,t') \, 
M^{\ast}_{\alpha_2 \gamma_2} (t,t') \, 
M_{\alpha_3 \gamma_3} (t,t')  
  \right) \; \; \; .
\end{eqnarray} 

\indent{In} summary, we have presented in this paper a framework to
study real-time evolution of scalar field theory. The technique has 
been applied to nuclear many-body theory and extended recently in the
context of homogeneous field configurations. Here we discuss this
general problem when the spatial dependence is important. Here we show
that the spatial dependence can be treated in the general orbital
representation and the dynamics of these orbitals are expressed in
a closed form by a set of selfconsistent equations. In this way, the
time-dependent technique can be used to improve the usual
gaussian-like mean field approximation, where the collisional dynamics
are given by approprite memorial integrals. We have illustrated these
procedures within the simplest context of self-interacting $\phi^4$
theory. 

\bigskip
\renewcommand{\theequation}{A.\arabic{equation}}
\setcounter{equation}{0}
\begin{center}
{\bf Appendix A: Projection Technique in Field Theory}
\end{center}

In order to calculate equations of motion (2.24), (2.32), 
(2.33) and (2.34) we first decompose \ $F$ \  in two parts
\begin{equation}
F \;  = \; F_0(t) + F'(t) 
\end{equation} 
where \ $F_0(t)$ \  is  the exponential of  a one-boson density given
in (3.39). 
A  crucial point  is to observe that \ $F_0(t)$ \ can 
be seen as a time-dependent projection of \ $F \, $, i.e.,
For the explicit construction of \ $\PP (t)$ \ we require, in addition
to eqs.~(3.3), the condition 
\begin{equation}
i \, \dot{F}_0 (t) \;  = \;  [F_0 (t) \,  , \, H] \, + \, \PP (t)[H,F] 
\end{equation} 
which is the Heisenberg  picture counterpart  of the Schr\"odinger
picture condition used in 
ref\cite{BFN88} 
to determine \ $\PP$ \ uniquely. \ 
The resulting form for  \ $\PP (t)$ \  is 
(see ref.\cite{LTP92} for details of the derivation)

\vspace{0.3cm}
\begin{eqnarray}
\PP \,  \cdot & = & \left\{ \left
[ 1 - \sum_{\alpha} \; \frac{d_{\alpha}^{\dag} \, d_{\alpha} -
p_{\alpha}}{1 + p_{\alpha}}  \right] \, {\rm Tr} (\cdot) \, + \,
\sum_{\alpha_1 \alpha_2} \; \frac{d^{\dag}_{\alpha_1} \, d_{\alpha_2}
 - p_{\alpha_2} \, \delta_{\alpha_{1} \alpha_{2}}}{p_{\alpha_{2}} (1 +
p_{\alpha_{1}})} \; \,  {\rm Tr} \left( d^{\dag}_{\alpha_{2}} \,
d_{\alpha_{1}} \, \cdot \right)  \right. \nonumber \\ 
\nonumber \\ 
& + & \sum_\alpha \left[ \,  \frac{d_{\alpha}}{p_{\alpha}} \; \,  {\rm Tr} \left( d_{\alpha}^{\dag} \, \cdot \right) \; + \; \frac{d^{\dag}_{\alpha}}{1 + p_{\alpha}} \; \, {\rm Tr} (d_{\alpha} \, \cdot) \,  \right] \;  + \; \sum_{\alpha_1 \alpha_2} \,  \left[ \, \frac{d_{\alpha_1} \, d_{\alpha_2}}{2 p_{\alpha_1} \, p_{\alpha_2}} \,  \; {\rm Tr} \left( d^{\dag}_{\alpha_2} \, d^{\dag}_{\alpha_1} \, \cdot \right)   \right. \nonumber \\ 
\nonumber \\ 
& + & \left. \left. \frac{d^{\dag}_{\alpha_1} \, d^{\dag}_{\alpha_2}}{2(1 + p_{\alpha_1})(1 + p_{\alpha_2})} \, \; {\rm Tr} \left( d_{\alpha_2} \, d_{\alpha_1} \, \cdot \right) \,  \right]  \right\} \, F_0 \; \; \; .
\end{eqnarray} 

\vspace{0.3cm}

The  next  step  is  to obtain  a  differential  equation of \ $F'(t)$. \   
This follows immediately from eqs.~(3.1) and (3.4),
\vspace{0.3cm}
\begin{equation}
\left( i \; \frac{d}{dt} \,  - \, \PP (t) \, \LL \right) \, F'(t) \; \,  = \, \; \Q \; (t) \; \LL \,  F_0(t) \; \; \; ,
\end{equation} 

\vspace{0.3cm}

\noindent{where} we introduced the operators
\begin{equation}
\Q \; (t) \; = \; \LL - \PP (t) \; \; \; \; \; , \; \; \; \; \; \LL \, \cdot \; = \; [H \, , \, \cdot] \; \; \; .
\end{equation} 
This equation has the formal solution
\vspace{0.3cm}
\begin{equation}
F'(t) \; = \; \GG \, \; (t,0) \, F'(0) \, - \, i \int^t_0  dt' \, \GG \, \; (t,t') \, \Q \; (t') \,  \LL  \, F_0(t') \; \; \; ,
\end{equation} 

\vspace{0.3cm}

\noindent{where}  \ $\GG \, \; (t,t')$ \  is the time-ordered Green's Function
\vspace{0.3cm}
\begin{equation}
\GG \, \; (t,t') \; \,  = \, \; T \; \exp \, i \int^t_{t'} d \tau \, \PP (\tau) \, L  \; \; \; .
\end{equation} 

\vspace{0.3cm}

A systematic expansion to treat the memory integral of the equation~(3.8) has been 
discussed in 
ref.\cite{BFN88}
in the Schr\"odinger picture. \ The implementation 
of the corresponding expansion in the Heisenberg picture consists 
in approximating the time evolution of the field operators by the simpler mean-field Hamiltonian 
\vspace{0.3cm}
\begin{eqnarray}
H_0 & = & P^{\dag} \, H + \sum_{\alpha} \, d^{\dag}_{\alpha} \; {\rm Tr}[d_{\alpha} \, , \, H] \, F' - \sum_{\alpha} \,  d_{\alpha} \; {\rm Tr} \left[ d^{\dag}_{\alpha} \, , \, H \right] \, F' \nonumber \\ 
\nonumber \\
& & + \;  \sum_{\alpha_1 \alpha_2} \; \frac{d^{\dag}_{\alpha_1} \, d^{\dag}_{\alpha_2}}{2(1 + p_{\alpha_1} + p_{\alpha_2})} \; \, {\rm Tr}[d_{\alpha_1} \, d_{\alpha_2} \, , \, H] \, F' \nonumber \\ 
\nonumber \\ 
& &  - \; \sum_{\alpha_1 \alpha_2} \; \frac{d_{\alpha_1} \, d_{\alpha_2}}{2(1 + p_{\alpha_1} + p_{\alpha_2})} \; \, {\rm Tr} \left[d^{\dag}_{\alpha_1} \, d^{\dag}_{\alpha_2} \, , \, H \right] \, F' .  \; \; \; .
\end{eqnarray} 

\vspace{0.3cm}

\noindent{Using} this approximation, one can solve explicitly the 
Heisenberg field operator equation as
\vspace{0.3cm}
\begin{equation}
d_{\alpha} (t) \; \; \simeq \; \; \sum_{\gamma} \; M_{\alpha \gamma} (t,t') \; d_{\gamma} (t') \; \; \; ,  
\end{equation} 

\vspace{0.3cm}

\noindent{where} the matrix \ $M_{\alpha \gamma}$ \ is the solution of the matrix equation
\begin{equation}
i \, \dot{M} (t,t') \; = \; F(t) \, M(t, t') \; \; \; ,
\end{equation} 
and the matrix \ $F$ \ involves matrix \ $h$ \ and mean-field 
energy (see appendix~A for details of the derivation).

In this way, the authors of the 
ref.\cite{BFN88} 
devised a systematic 
expansion for the correlation density \ $F'(t)$. \  The corresponding 
expansion in the Heisenberg picture is 
\vspace{0.3cm}
\begin{eqnarray}
F'(t) & = & \GG \, \;  (t,0) \, F'(0) \, - \, i \int^t_0  dt'_1 \, Q (t_1) \, L  \, F_0 (t_1)  \nonumber \\ 
\nonumber \\ 
& - &  \int^t_0 dt_1 \, \left[ \, \int^t_{t_1} dt_2 \, Q(t_2) \, (L - L_0 (t_2) \right] \, Q(t_1) \, L \,  F_0 (t_1) + \cdots  \; \; \;  ,
\end{eqnarray} 

\vspace{0.3cm}

\noindent{where} \ $L_0 \, \cdot = [H_0 \, , \, \cdot]$. \ In what follows we restrict 
ourselves to initial conditions such that \ $F'(0) = 0$ \ and to the 
lowest approximation for \ $F'(t)$.  \ Therefore, the evaluation of the equations of motion involves traces of the type
\vspace{0.3cm}
\begin{equation}
{\rm Tr} \, [ \hat{O} (t), H] \, F_0(t) \; - \; i \; {\rm Tr} \, [ \hat{O} (t), H] 
\int^t_0 dt' \, \Q \; (t') \, [H, F_0(t') ] \; \; \; , 
\end{equation} 

\vspace{0.3cm}

\noindent{where} \ $\hat{O} (t)$ \ can be \ $d_{\alpha} (t) \, , \; d^{\dag}_{\alpha} (t) \, d_{\beta} (t) \, , \; d_{\alpha} (t) \, d_{\beta} (t) \, $, \ and 
operators at different times are related by equation~(3.11).

\renewcommand{\theequation}{B.\arabic{equation}}
\setcounter{equation}{0}
\begin{center}
{\bf Appendix B: \  Approximation for the Time Evolution of \\ the
  Heisenberg  Field Operator} 
\end{center}

\vspace{0.3cm}

In section~II, we have discussed that our approximation consists in replacing 
the time evolution of the field operator by a simpler mean-field Hamiltonian 
given by equation~(3.10). \ We show now that this allows one to solve the 
Heisenberg operator equation 
\vspace{0.3cm}
\begin{equation}
i \, \dot{d}_{\alpha} \; = \; [d_\alpha , H_0]  - \sum_\gamma \, h_{\alpha \gamma} \, d_{\alpha} - 
 \sum_\gamma \, g_{\gamma \alpha} \, d^{\dag}_{\gamma} - {\rm Tr} \, [d_\alpha , H]  \; \; \; .
\end{equation} 

\vspace{0.3cm}

\noindent{The} last three terms are the explicit time dependence of \ $d_\alpha (t)$ \ 
related to shifted amplitudes \ $A_k (t)$ \ and to effects of the general Bogoliubov transformation for operators (2.19).

First, we write \ $P^{\dag} H \, $, \ using ciclyc properties of the traces, 
as (see ref.~[ \ \ \ \ \ ] for the construction of \ $P^{\dag}$)
\vspace{0.3cm} 
\begin{eqnarray}
P^{\dag} H & = & 
\left( 1 \, - \, \sum_\alpha \; \, \frac{d^{\dag}_{\alpha} \ d_\alpha - p_\alpha}{1 + p_\alpha}  \right) \, {\rm Tr} \, (H \, F_0) \, + \, 
\sum_{\alpha_1 \alpha_2} \; \, \frac{d^{\dag}_{\alpha_1} \ d_{\alpha_2} - p_{\alpha_1} \, \delta_{\alpha_1 \alpha_2}}{P_{\alpha_2} (1 + p_{\alpha_1})}  \; {\rm Tr} \, (d^{\dag}_{\alpha_2} \, d_{\alpha_1} \, H \, F_0) \nonumber \\
\nonumber \\
& - & \sum_{\alpha} \, d_{\alpha} \, {\rm Tr} \, [ d^{\dag}_{\alpha} , H] \, F_0 \, + \, 
\sum_{\alpha} \, d^{\dag}_{\alpha} \, {\rm Tr} \, [ d_{\alpha} , H] \, F_0  \nonumber \\
\nonumber \\
& - & \sum_{\alpha_1 \alpha_2} \; \, \frac{d_{\alpha_1} \, d_{\alpha_2}}{2 (1 + p_{\alpha_1} + p_{\alpha_2})} \, \; {\rm Tr} \, [d^{\dag}_{\alpha_2} \, d^{\dag}_{\alpha_2} \, , \, H] \, F_0 \nonumber \\
\nonumber \\
& + & \sum_{\alpha_1 \alpha_2} \; \, \frac{d^{\dag}_{\alpha_1} \, d^{\dag}_{\alpha_2}}{2 (1 + p_{\alpha_1} + p_{\alpha_2})} \, \; {\rm Tr} \, [d_{\alpha_2} \, d_{\alpha_1} \, , \, H] \, F_0
\end{eqnarray} 

\vspace{0.3cm}

\noindent{Hence}, equation (3.10) becomes
\vspace{0.3cm}
\begin{eqnarray}
H_0 & = & 
\left( 1 \, - \, \sum_\alpha \; \, \frac{d^{\dag}_{\alpha} \ d_\alpha - p_\alpha}{1 + p_\alpha}  \right) \, {\rm Tr} \, (H \, F_0) \, + \, 
\sum_{\alpha_1 \alpha_2} \; \, \frac{d^{\dag}_{\alpha_1} \ d_{\alpha_2} - p_{\alpha_1} \, \delta_{\alpha_1 \alpha_2}}{p_{\alpha_2} (1 + p_{\alpha_1})}  \; {\rm Tr} \, (d^{\dag}_{\alpha_2} \, d_{\alpha_1} \, H \, F_0) \nonumber \\
\nonumber \\
& - & \sum_{\alpha} \, d_{\alpha} \, {\rm Tr} \, [ d^{\dag}_{\alpha} , H] \, F \, + \, 
\sum_{\alpha} \, d^{\dag}_{\alpha} \, {\rm Tr} \, [ d_{\alpha} , H] \, F  \nonumber \\
\nonumber \\
& - & \sum_{\alpha_1 \alpha_2} \; \, \frac{d_{\alpha_1} \, d_{\alpha_2}}{2 (1 + p_{\alpha_1} + p_{\alpha_2})} \, \; {\rm Tr} \, [d^{\dag}_{\alpha_2} \, d^{\dag}_{\alpha_2} \, , \, H] \, F  \nonumber \\ 
\nonumber \\ 
& + & \sum_{\alpha_1 \alpha_2} \; \, \frac{d^{\dag}_{\alpha_1} \, d^{\dag}_{\alpha_2}}{2 (1 + p_{\alpha_1} + p_{\alpha_2})} \, \; {\rm Tr} \, [d_{\alpha_2} \, d_{\alpha_1} \, , \, H] \, F
\end{eqnarray} 

\vspace{0.3cm}

\noindent{Using} (A.3) in (A.1) one obtains immediately
\vspace{0.3cm}
\begin{equation}
i \, \dot{d}_{\alpha} \; = \; 
- \; \frac{{\rm Tr} (HF_0}{1 + p_{\alpha}} \; d_{\alpha} + \sum_{alpha'} \; \frac{d_{\alpha'}}{p_{\alpha'}(1 + p_{\alpha})} \; {\rm Tr} (d^{\dag}_{\alpha'} \, d_{\alpha} \, H \, F_0) - \sum_{\alpha} \, h_{\alpha \gamma} \, d_{\gamma} \; \; \; .
\end{equation} 

\vspace{0.3cm}

\noindent{The} calculation for the first two terms is straightforward. \  It yields 
\vspace{0.3cm}
\begin{eqnarray}
&  & \frac{1}{2} \; \sum_{\alpha'} \,  \sum_k \, \left[ 
k_0 \, (U^{\ast}_{k \alpha} + V^{\ast}_{- k \alpha} ) (U_{k \alpha'} + V_{- k \alpha'} ) + 
\frac{k^2 + m^2}{k_0} \; \, 
T^{\ast}_{k \alpha} \,  T_{k \alpha'}  \right] \, d_{\alpha'} \nonumber \\ 
\nonumber \\ 
& & + \;  \frac{g}{16L} \;  \sum_{k_1 k_2 k_3 k_4} \; \, \frac{1}{\sqrt{k_{01} \, k_{02} \, k_{03} \, k_{04}}} \; \, \sum_{\alpha'} \nonumber \\ 
\nonumber \\ 
&  & \times \; \left\{  \delta_{k_1 + k_2 - k_3 - k_4 \, ,\, 0} \; \left( A_{k_1} \, A_{k_2}  \, T^{\ast}_{k_3 \alpha} \, T_{k_4 \alpha'} \, + \, A^{\ast}_{k_1} \, A^{\ast}_{k_2}  \, T_{k_3 \alpha'} \, T_{k_4 \alpha} \right) \right.  \nonumber \\
\nonumber \\
&  & + \; \delta_{k_1 + k_2 + k_3 - k_4 \, ,\, 0} \; \left( A_{k_1} \, A_{k_2}  \, T_{k_3 \alpha'} \, T^{\ast}_{k_4 \alpha} \, + \, A^{\ast}_{k_1} \, A^{\ast}_{k_2}  \, T^{\ast}_{k_3 \alpha} \, T_{k_4 \alpha'} \right)  \nonumber \\
\nonumber \\
&  & + \;  \left. 2 \delta_{k_1 - k_2 + k_3 - k_4 \, ,\, 0} \; \left( A^{\ast}_{k_1} \, A_{k_2}  \, T^{\ast}_{k_3 \alpha} \, T_{k_4 \alpha'} \, + \, A_{k_1} \, A^{\ast}_{k_2}  \, T_{k_3 \alpha'} \, T^{\ast}_{k_4 \alpha} \right) \right \} \, d_{\alpha'} \nonumber \\
\nonumber \\
&  & + \;  \frac{g}{4L} \;  \sum_{k_1 k_2 k_3 k_4} \; \frac{\delta_{k_1 + k_2 - k_3 - k_4 \, ,\, 0}}{\sqrt{k_{01} \, k_{02} \, k_{03} \, k_{04}}} \; \sum_{\alpha_1 \alpha_2} \; 
 T^{\ast}_{k_1 \, \alpha} \, T^{\ast}_{k_2 \, \alpha_2} \,  T_{k_3 \, \alpha_1} \, T_{k_4 \, \alpha_2} \, p_{\alpha_2} \, d_{\alpha_1}  \nonumber \\ 
\nonumber \\ 
&  & + \;  \frac{g}{8L} \;  \sum_{k_1 k_2 k_3 k_4} \; \frac{\delta_{k_1 + k_2 - k_3 - k_4 \, ,\, 0}}{\sqrt{k_{01} \, k_{02} \, k_{03} \, k_{04}}} \; \sum_{\alpha_1 \alpha_2} \; 
 T^{\ast}_{k_1 \, \alpha_1} \, T^{\ast}_{k_2 \, \alpha_2} \,  T_{k_3 \, \alpha_2} \, T_{k_4 \, \alpha_1} \, d_{\alpha_1} \, d_{\alpha_1} ) \nonumber \\ 
\nonumber \\
&  & - \;  \frac{g}{8L} \;  \sum_{k_1 k_2} \; \frac{1}{\sqrt{k^2_1 + m^2}} \; \, \frac{1}{k_{02}} 
\; \sum_{\alpha_1} \; 
T^{\ast}_{k_2 \, \alpha} \,  T_{k_2 \, \alpha_1} \, d_{\alpha_1} \nonumber \\ 
\nonumber \\
& & \equiv \; \sum_{\gamma} \; B_{\alpha \gamma} \, d_{\gamma} \; \; \; .
\end{eqnarray} 

\vspace{0.3cm}

One finds finally the Heisenberg operator equation as being
\vspace{0.3cm}
\begin{eqnarray}
i \, \dot{d}_{\alpha} & = & \sum_{\gamma} \; B_{\alpha \gamma} \, d_{\gamma} \; - \; 
\sum_{\gamma} \; h_{\alpha \gamma} \, d_{\gamma} \nonumber \\
\nonumber \\
& \equiv & \sum_{\gamma} \; F_{\alpha \gamma} \, d_{\gamma}
\end{eqnarray} 

\vspace{0.3cm}

\noindent{or in the matrix form}
\begin{equation}
i \, \dot{d} (t) \; \; = \; \; F(t) \, d(t) \; \; \; .
\end{equation} 
It is easy to see that the solution for (A.7) is
\begin{equation}
d(t) \; \; = \; \; M(t,t') \, d(t') \; \; \; , 
\end{equation} 
where \ $M(t,t')$ \ is the solution of the matrix equation 
\begin{equation}
i \, \dot{M} (t,t') \; \; = \; \; F(t) \, M(t,t') 
\end{equation} 
with initial condition
\begin{equation}
M(t,t') \; \; = \; \; 1 \; \; \; \; .
\end{equation} 

\end{document}